\documentclass[]{elsart}  
\usepackage[dvips]{graphicx}

\def\micron{{$\mu$m}}
\def\degC{{${}^\circ$C}}
\def\degree{{${}^\circ$}}

\begin{document} 
\thispagestyle{empty}
\renewcommand{\thefootnote}{\fnsymbol{footnote}}

%%%%% Substitute your Pub number, month and year in the following:
%%
\begin{flushright}
{\small
SLAC--PUB--9638\\
January, 2003\\}
\end{flushright}

\vspace{.8cm}

%%%%% Title and Author Information:
%%
\begin{center}
{\bf\large   
Gamma-ray Polarimetry
\footnote{This work was carried out in collaboration with T.~Kamae and E. do~Couto~e~Silva (SLAC), 
S.~Uno, T.~Nakamoto and Y.~Fukazawa (Hiroshima University), 
T.~Mitani,  T.~Takahashi and K.~Nakazawa (Institute of Scape and Astronautical Science)
and D.~Marlow (Princeton University), 
and supported by U.S. Department of Energy, contract DE-AC03-76SF00515,
Grantin-Aid by Ministry of Education, Culture, Sports, Science and Technology of Japan (12554006, 13304014),
and ``Ground-based Research Announcement for Space Utilization'' promoted by Japan Space Forum.
}
}

\vspace{1cm}

Hiroyasu Tajima\\
Stanford Linear Accelerator Center, Stanford University,
Stanford, CA  94309\\

\end{center}

\vfill

\begin{center}
{\bf\large   
Abstract }
\end{center}

\begin{quote}
An astrophysics application of a low noise Double-sided Silicon Strip
Detector (DSSD) is described.  A Semiconductor Multiple-Compton
Telescope (SMCT) is being developed to explore the gamma-ray universe
in the 0.1--20~MeV energy band.  
Excellent energy resolution and polarization sensitivity are key features of the SMCT.  
We have developed prototype modules for a low noise DSSD system, which reached
an energy resolution of 1.3~keV (FWHM) for 122~keV at
0\degC.  Results of a gamma-ray imaging test are also presented.

%>>>> Include a list of keywords after the abstract 
%\begin{keyword}

\bigskip
\medskip

\noindent
{\it Keywords: gamma-ray, Compton telescope, polarimeter, Silicon Strip Detector}
%\end{keyword}

\end{quote}

\vfill

%%%%%%%%%%%%%%%
%% Choose"Presented at," "Contributed to" for conference papers
%% or "Submitted to" for journal papers
%%%%%%%%%%%%%%%
\begin{center} 
{\it Contributed to} 
{\it International Workshop on  Vertex Detectors}\\
{\it Kailua-Kona, Hawaii USA}\\
{\it November 3--November 8, 2002} \\

%%OR\\

%%{\it Submitted to Nuclear Instruments and Method}

\end{center}

\newpage

%%%%%%%%%%%%%%%%%%%%%%%%%%%%%%%%%%%%%%%%%%%%%%%%%%%%%%%%%%%%%
% Introduction
%%%%%%%%%%%%%%%%%%%%%%%%%%%%%%%%%%%%%%%%%%%%%%%%%%%%%%%%%%%%%
\section{Introduction}
%\label{sect:intro}  % \label{} allows reference to this section

A low noise Double-sided Silicon Strip Detector (DSSD) is a
fundamental element to realize a Semiconductor Multiple-Compton
Telescope (SMCT)\cite{Tajima02,Takahashi02}, which utilizes a
multiple-Compton scattering technique to achieve good efficiency with high
background rejection capability.  
The SMCT is a hybrid semiconductor gamma-ray detector, which consists of silicon and CdTe detectors.
The excellent energy resolution of the SMCT allows it to detect photons with
good angular resolution and background rejection capability.  
In addition, the ability to measure polarization, a wide energy band
(0.1--20 MeV), a wide field of view ($\sim$60\degree) and light weight
are important properties of the SMCT.  
Polarization sensitivity enables us to study particle acceleration mechanisms in supernovae
remnants, pulsars and back holes.  
Synchrotron radiation, inverse Compton scattering and bremsstrahlung are main photon processes and they produce distinct photon polarization features.

%%%%%%%%%%%%%%%%%%%%%%%%%%%%%%%%%%%%%%%%%%%%%%%%%%%%%%%%%%%%%
% Multi-Compton Technique.
%%%%%%%%%%%%%%%%%%%%%%%%%%%%%%%%%%%%%%%%%%%%%%%%%%%%%%%%%%%%%
%\section{Multiple-Compton Technique}

A stack of many thin scatterers is used in the multiple-Compton
technique\cite{Kamae87}, which allows increased detection
efficiency by stacking more elements, while maintaining the ability to
record individual Compton scatterings.  

%-------------
   \begin{figure}[tbh]
   \begin{center}
   \begin{tabular}{lll}
   (a) & (b) & (c) \\
\raisebox{0.5cm}[0pt]{\includegraphics[height=3.5cm]{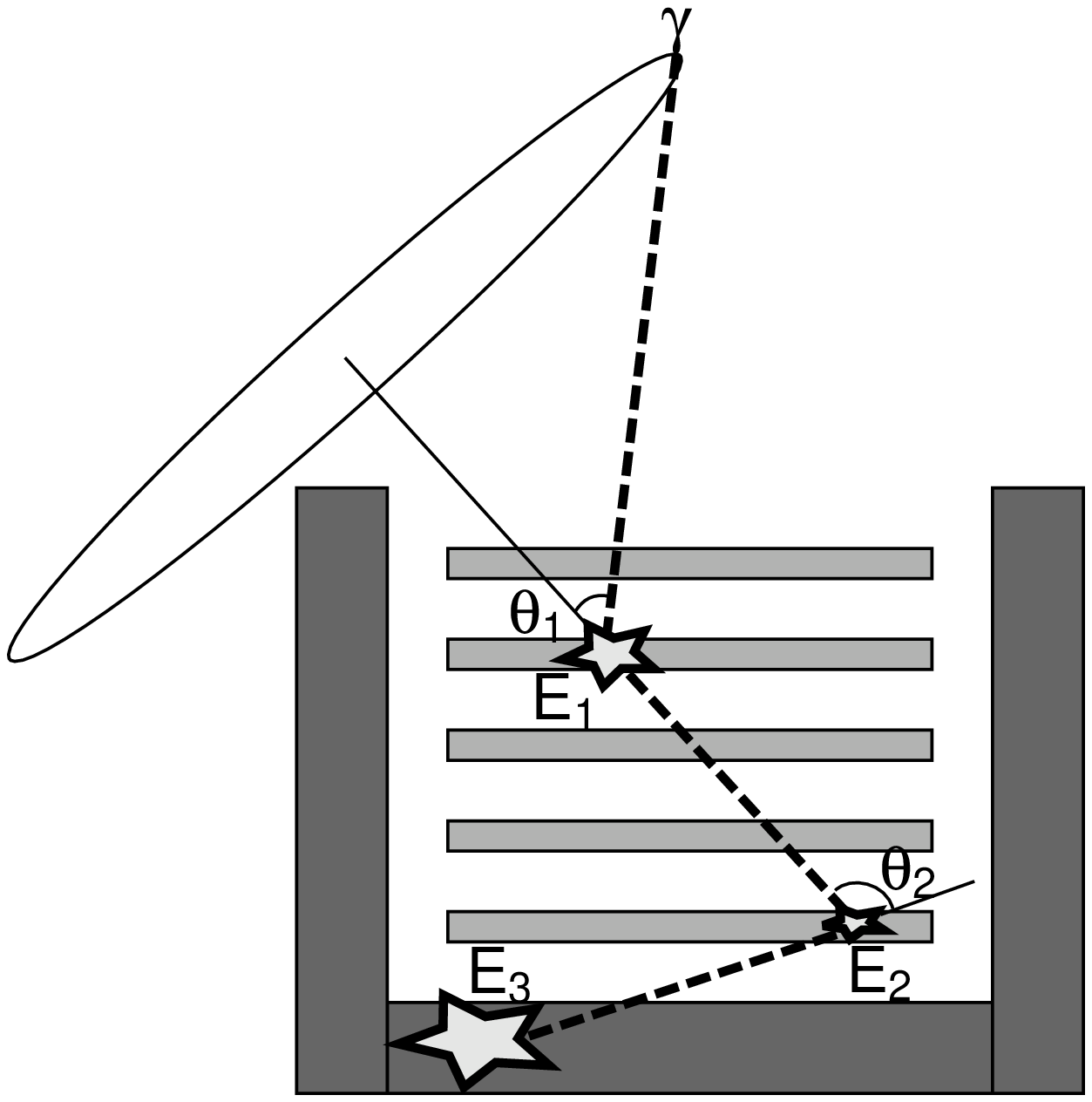} \hspace*{0.0cm}} 
	& \includegraphics[height=4cm]{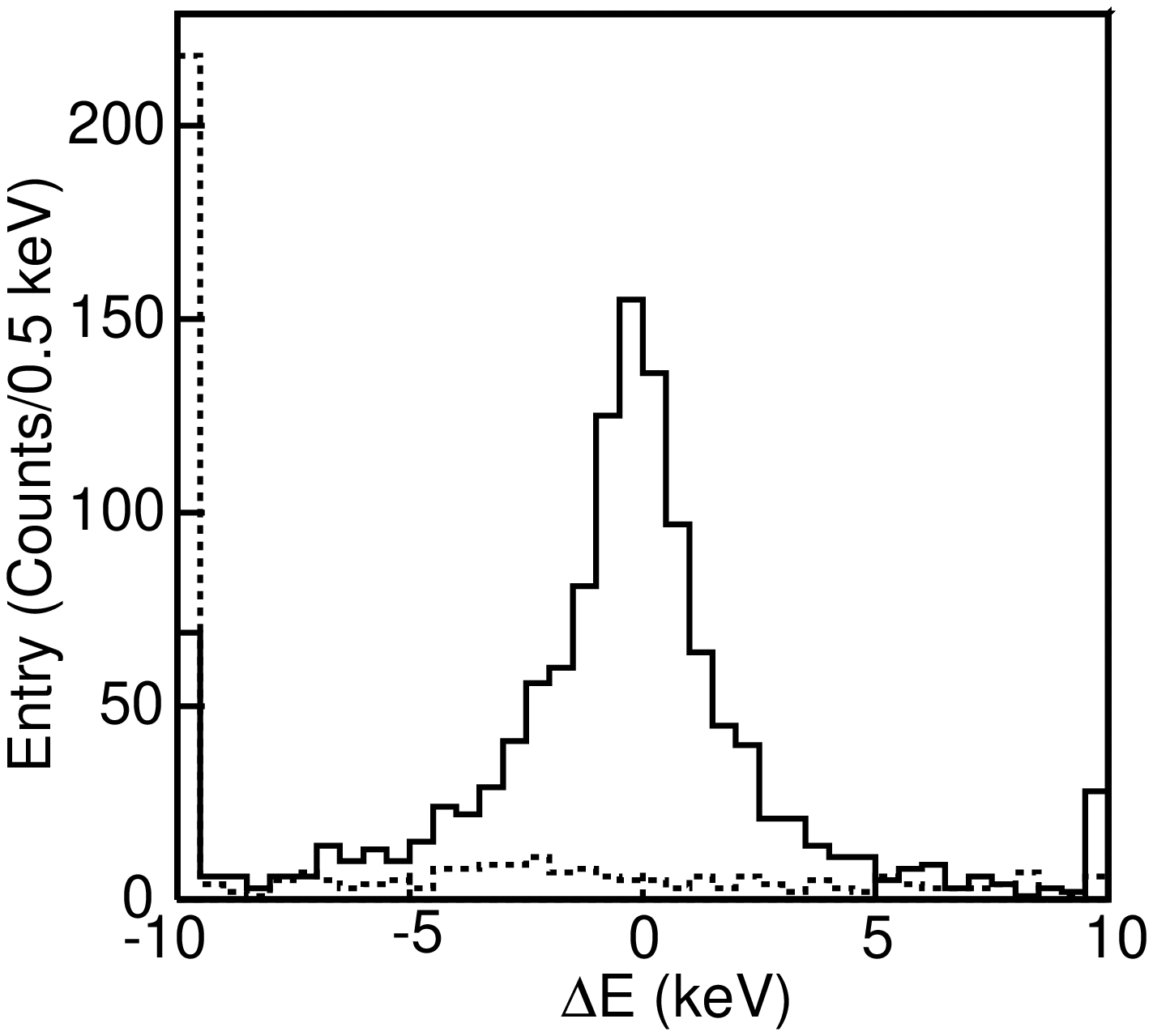}
    & \includegraphics[height=4cm]{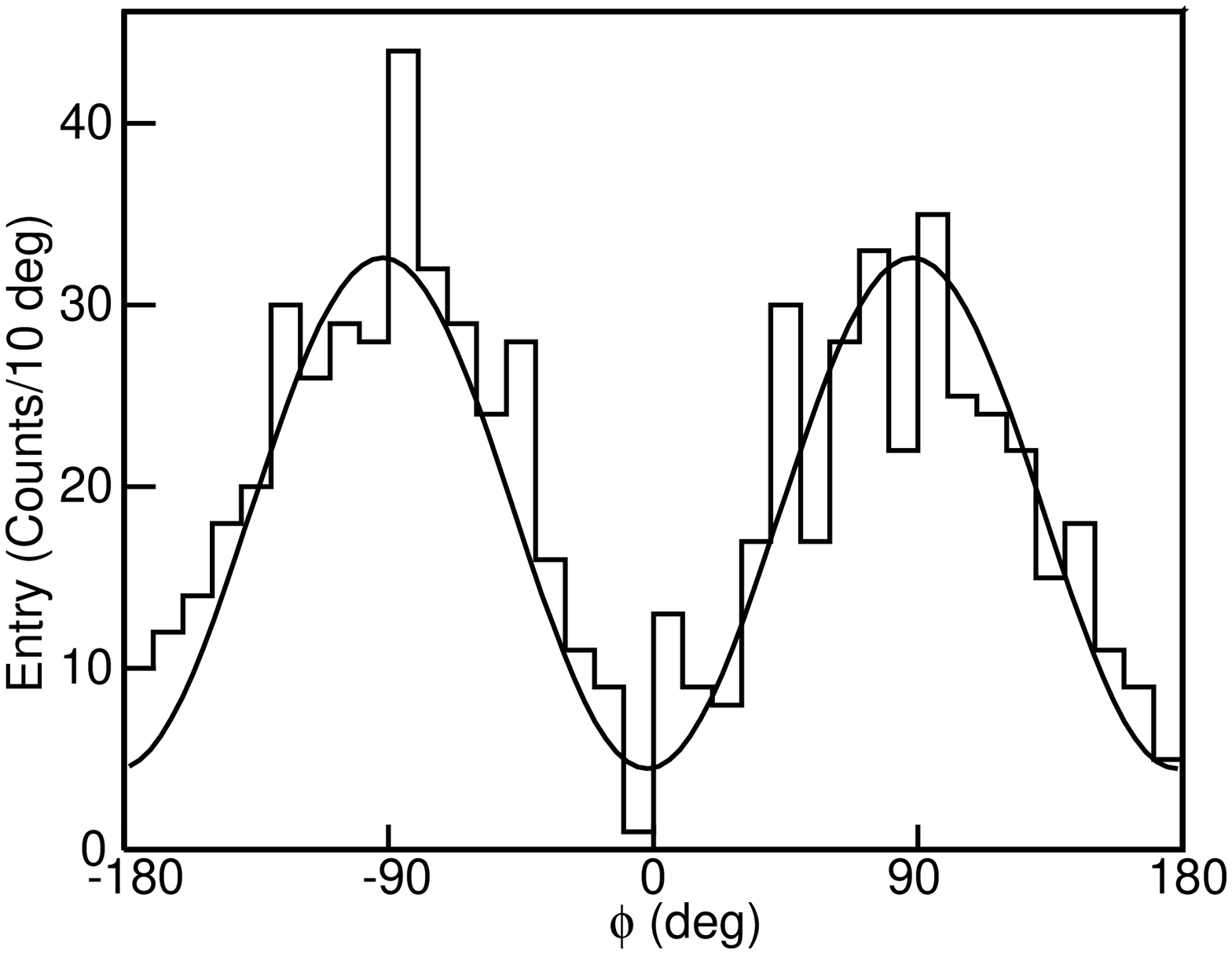}\\
   \end{tabular}
   \end{center}
   \caption[Concept of Multiple-Compton technique.] 
%>>>> use \label inside caption to get Fig. number with \ref{}
   { \label{fig:schematic} (a) Concept of the Multiple-Compton technique. 
(b) Difference between calculated and measured electron recoil energy inEGS4 MC. Solid and dotted histograms indicate signal and backgrounds, respectively.
(c) Reconstructed $\phi$ distribution of the first Compton scatter in MC with 100\% linear polarization.}
   \end{figure} 
%-------------
Fig.~\ref{fig:schematic} (a) illustrates the case with two Compton scatterings and one photoelectric absorption.  
In this situation, scattering angles $\theta_1$ and $\theta_2$ can both be obtained from the recoil electron energies as
%\begin{displaymath}
\begin{equation}
\cos\theta_1 = 1+\frac{m_ec^2}{E_1+E_2+E_3}-\frac{m_ec^2}{E_2+E_3},\ \ 
\cos\theta_2 = 1+\frac{m_ec^2}{E_2+E_3}-\frac{m_ec^2}{E_3},\label{eq:kinematics}
\end{equation}
%\end{displaymath}
where $E_1$, $E_2$ and $E_3$ are the energy deposits in each photon interaction.  
Note that $\theta_2$ can be reconstructed from the hit positions of the three interactions.  
The order of the interaction sequences, hence the correct energy and direction of the incident photon, can be reconstructed by examination of this constraint for all possible sequences.  
This over-constraint also provides stringent suppression of random coincidence backgrounds.
The direction of the incident photon can be confined in a cone determined from $\theta_1$ and the first two interaction positions.
%This technique has proven to be a very powerful method to suppress backgrounds.
%% that is crucial to observe faint or diffuse gamma-ray sources. 
%% - I commented this out since this does not by itself provide a 
%% - significant improvement in angular resolution, but rather helps
%% - identify ``goodness of fit'', and so could be misleading as stated.
%%
Fig.~\ref{fig:schematic} (b) shows a distribution of the difference of
the recoil electron energy calculated from the scattering angle and the
``measured" energy from a EGS4 Monte Carlo simulation\cite{EGS4} with
low energy extension\cite{EGS-KEK}.  Solid and dotted histograms
indicate signals and backgrounds, respectively.  Underflows and
overflows are shown in the leftmost and rightmost bins.  
% This clearly demonstrates the usefulness of the constraints from Compton
% kinematics.  -- unnecessary
The polarization of the incident photon can be measured by the azimuthal
angle ($\phi$) distribution of the first Compton scatter.
Fig.~\ref{fig:schematic} (c) shows the $\phi$ distribution reconstructed
from hits in DSSDs in a MC sample for 100~keV photons with 100\%
linear polarization.  Clear modulation can be observed.

%%%%%%%%%%%%%%%%%%%%%%%%%%%%%%%%%%%%%%%%%%%%%%%%%%%%%%%%%%%%%
% low noise SSD
%%%%%%%%%%%%%%%%%%%%%%%%%%%%%%%%%%%%%%%%%%%%%%%%%%%%%%%%%%%%%
\section{Low Noise SSD System}

We have developed prototype modules for a low noise DSSD system in
order to understand all noise sources in detail, which is necessary
to achieve the best possible energy resolution.  
A low noise DSSD system consists of a DSSD, an RC chip and a VA32TA front-end VLSI
chip\cite{Tajima02,VA94}.
To keep the strip yield close to 100\%, the DSSD does not employ an integrated AC capacitor.  
We have produced 300~\micron\ thick DSSDs with strip lengths of 2.56~cm, strip gaps of 100~\micron\ and strip pitches of either 400~\micron\ or 800~\micron.
%, and strip gaps of 100, 130 or 160~\micron\ to optimize strip geometry for the best noise performance.  
%A smaller strip gap is preferred to maintain higher breakdown voltage and allow future usage of thicker sensors.
%Negligible improvement in noise performance is observed for larger strip gap.  
The C-V curve measurement yields the depletion voltage of 65~V, 
therefore the following measurements are performed at 70~V bias.  
%No junction breakdown was observed up to 200~V bias voltage.  
Leakage current is measured to be 0.5~nA/strip at 20\degC\ and 0.05~nA/strip
at 0\degC.  
The strip capacitance is measured to be $6.3\pm0.2$~pF for a 400~\micron\ pitch sample.
%The RC chip provides detector bias voltage via polysilicon bias resistors, as well as AC-coupling between strips and preamplifier channel inputs.  
A resistance value of 1 G$\Omega$ is chosen for bias resister to minimize thermal noise without compromising production stability.  
%The size of this MOS capacitor is 1.2~mm $\times$ 0.14~mm in order to
%obtain sufficient coupling capacitance while presenting a capacitance
%load to the preamplifier of less than 1~pF.
%We have developed the VA32TA front-end ASIC based on the VA32C amplifier VLSI and the TA32C trigger VLSI that are originally developed by Ideas ASA\cite{VA94}.  
%The VA32TA is fabricated in the AMS 0.35 $\mu$m CMOS process, which is measured to be radiation tolerant to 20~MRad or more\cite{Yokoyama01}.
%A VA32TA consists of 32 channels of signal-readout.  Each channel includes a charge sensitive preamplifier-slow CR-RC shaper-sample/hold-analog multiplexer chain (VA section) and fast shaper-discriminator chain (TA section).
%The front-end MOSFET size for the preamplifier is optimized for small capacitance load, 
%to attain optimal noise performance.
Expected noise performance of VA32TA is $(45+19\times C_d)/\sqrt{\tau}\;e^-$
(RMS) in ENC or $(0.37+0.16\times C_d)/\sqrt{\tau}$~keV (FWHM), where
$C_d$ is the detector load capacitance and $\tau$ is the peaking time, which
can be varied from 1 to 4 $\mu$s.

We have assembled two prototype modules: one consists of a 400 \micron\ pitch SSD, an RC chip and a VA32TA (AC module);  the other consists of an SSD and a VA32TA (DC module).  
%Noise performance of the prototype system is measured at temperatures of 0\degC\ and 20\degC\ and at peaking times of 2~$\mu$s and 4~$\mu$s.  
%Varying these parameters is useful to differentiate the noise contributions.  
Absolute gain of the system is calibrated using the $\gamma$-ray spectra obtained below.  
The DC module has yielded a best noise performance of 1.0~keV at 0\degC\
and $\tau = 4$~$\mu$s, which is in a good agreement with the expected value
derived from known noise performance of VA32TA and the measured strip
capacitance.  
%This indicates that the noise sources from the SSD and
%the VA32TA are well understood.

%-------------
   \begin{figure}[tbh]
   \begin{center}
   \begin{tabular}{ll}
(a) ${}^{241}$Am spectrum & (b) ${}^{57}$Co spectrum\\
   \includegraphics[height=4.1cm]{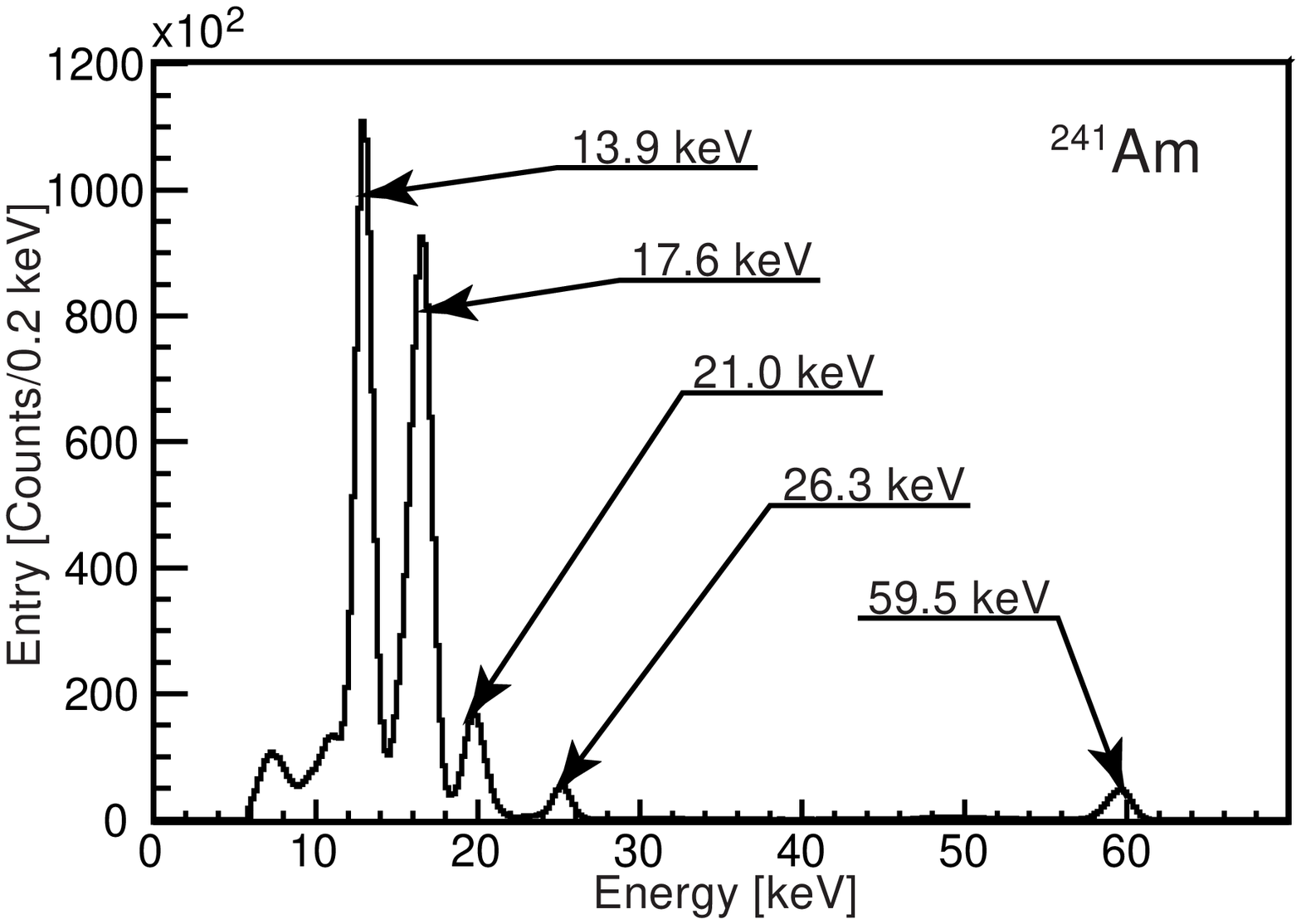} \hspace*{0.4cm} &
   \includegraphics[height=4.0cm]{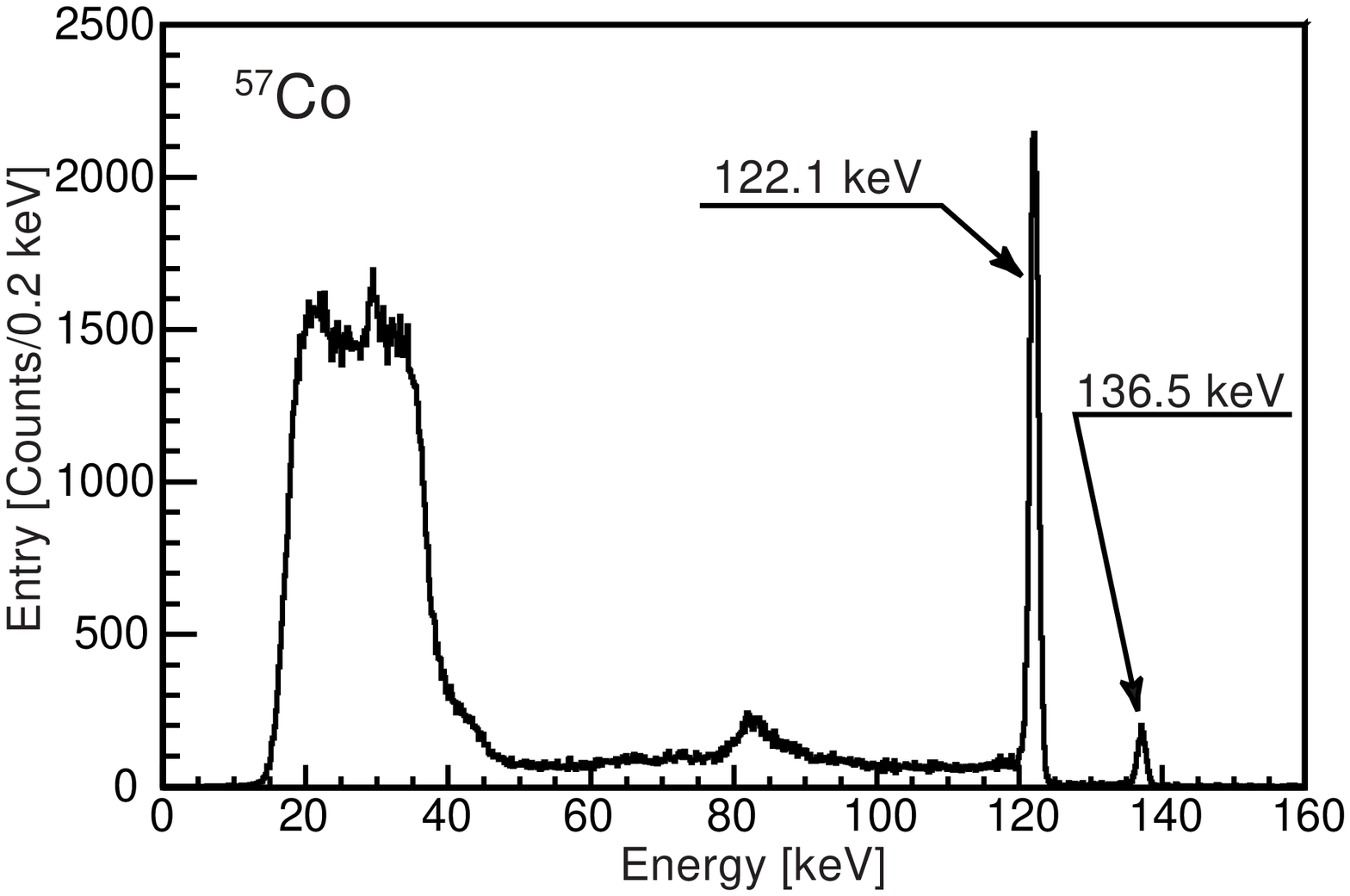}
   \end{tabular}
   \end{center}
   \caption
   { \label{fig:spectrum} Energy spectra of (a) ${}^{241}$Am and (b) ${}^{57}$Co. }
   \end{figure} 
%-------------
The energy resolution for the $\gamma$-ray detection is investigated
using the 59.54~keV $\gamma$-ray line from ${}^{241}$Am and the
122.06~keV $\gamma$-ray line from ${}^{57}$Co.  
Absolute gain is calibrated for each channel using the same $\gamma$-ray sources in
such way that all channels give the nominal peak height.
Figs.~\ref{fig:spectrum} (a) and (b) show the sum of the energy
spectra for all channels, except for the first and last strips where we observe larger noise.  
We observe a clear Compton edge just below 40~keV in the ${}^{57}$Co spectrum.  
%Peak positions other than 59.5~keV and 122.1~keV are somewhat shifted since a gain calibration has not yet been performed for these energy points.
Note only 59.5~keV and 122.1~keV lines are calibrated.
Energy resolution is measured to be 1.3~keV at 0\degC\ and 4~$\mu$s for both 59.54 and
122.06~keV $\gamma$-rays.  
A detailed description of the SSD system and the noise and energy resolution measurements can be found elsewhere\cite{Tajima02}.

%%%%%%%%%%%%%%%%%%%%%%%%%%%%%%%%%%%%%%%%%%%%%%%%%%%%%%%%%%%%%
% test results
%%%%%%%%%%%%%%%%%%%%%%%%%%%%%%%%%%%%%%%%%%%%%%%%%%%%%%%%%%%%%
\section{Imaging Test Results and Conclusions}
\begin{figure}[tbh]
   \begin{center}
   \begin{tabular}{ll}
(a) & (b)\\
  \hspace*{0.3cm} \includegraphics[height=4.0cm]{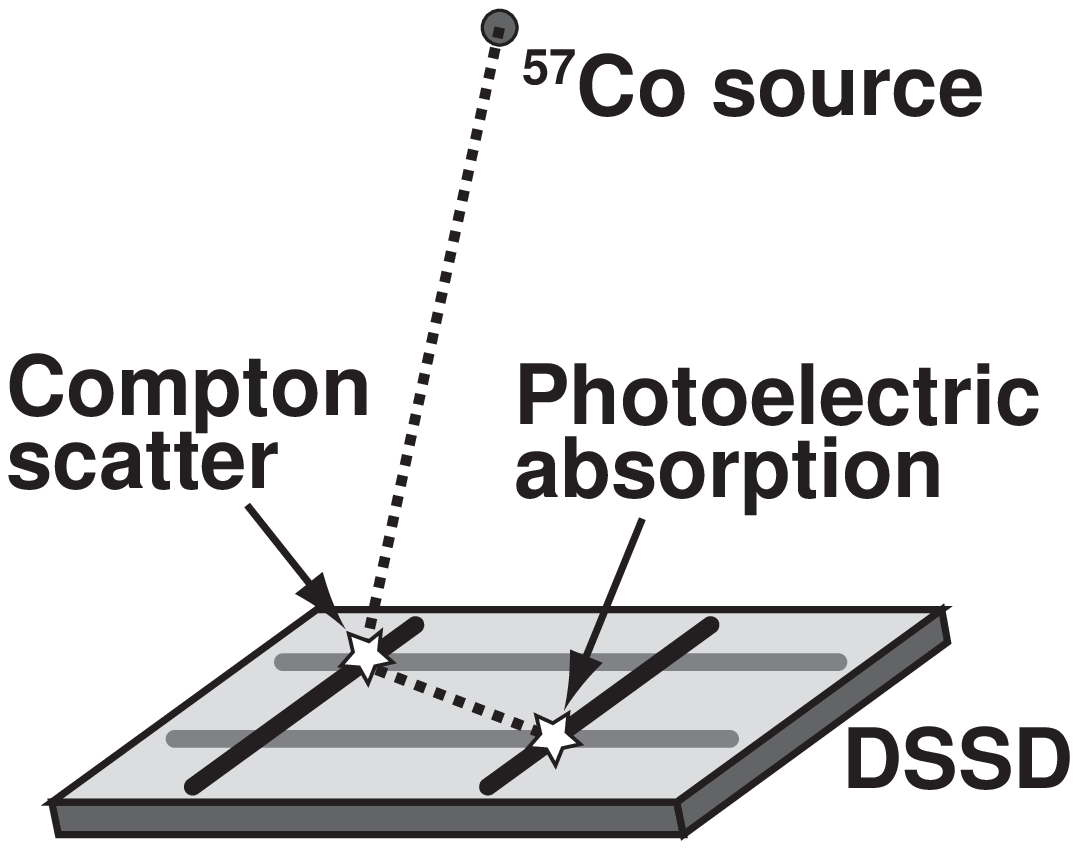} \hspace*{0.75cm} &
   \includegraphics[height=4.0cm]{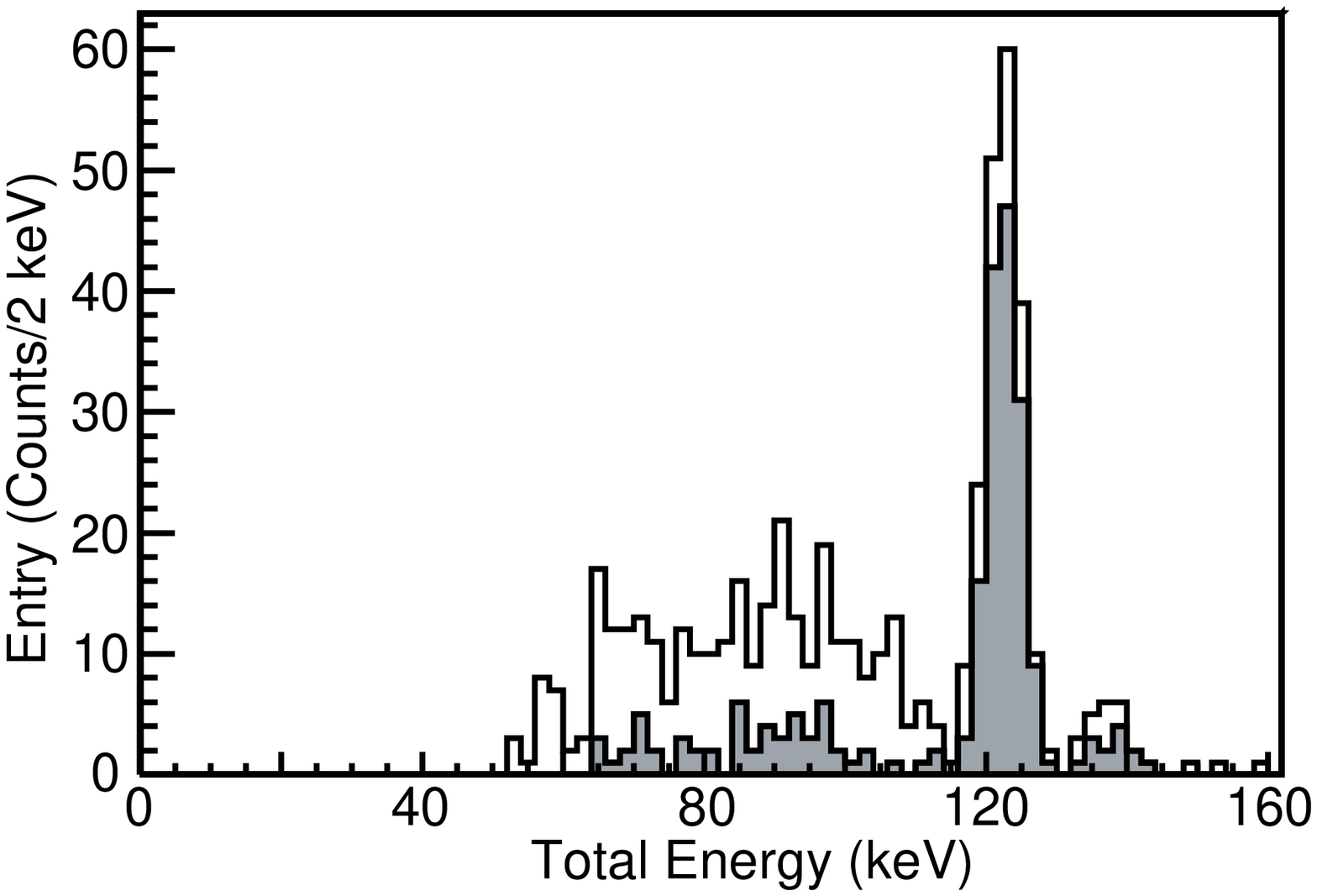}
   \end{tabular}
   \begin{tabular}{ll}
(c) & (d) \\
   \includegraphics[height=4.0cm]{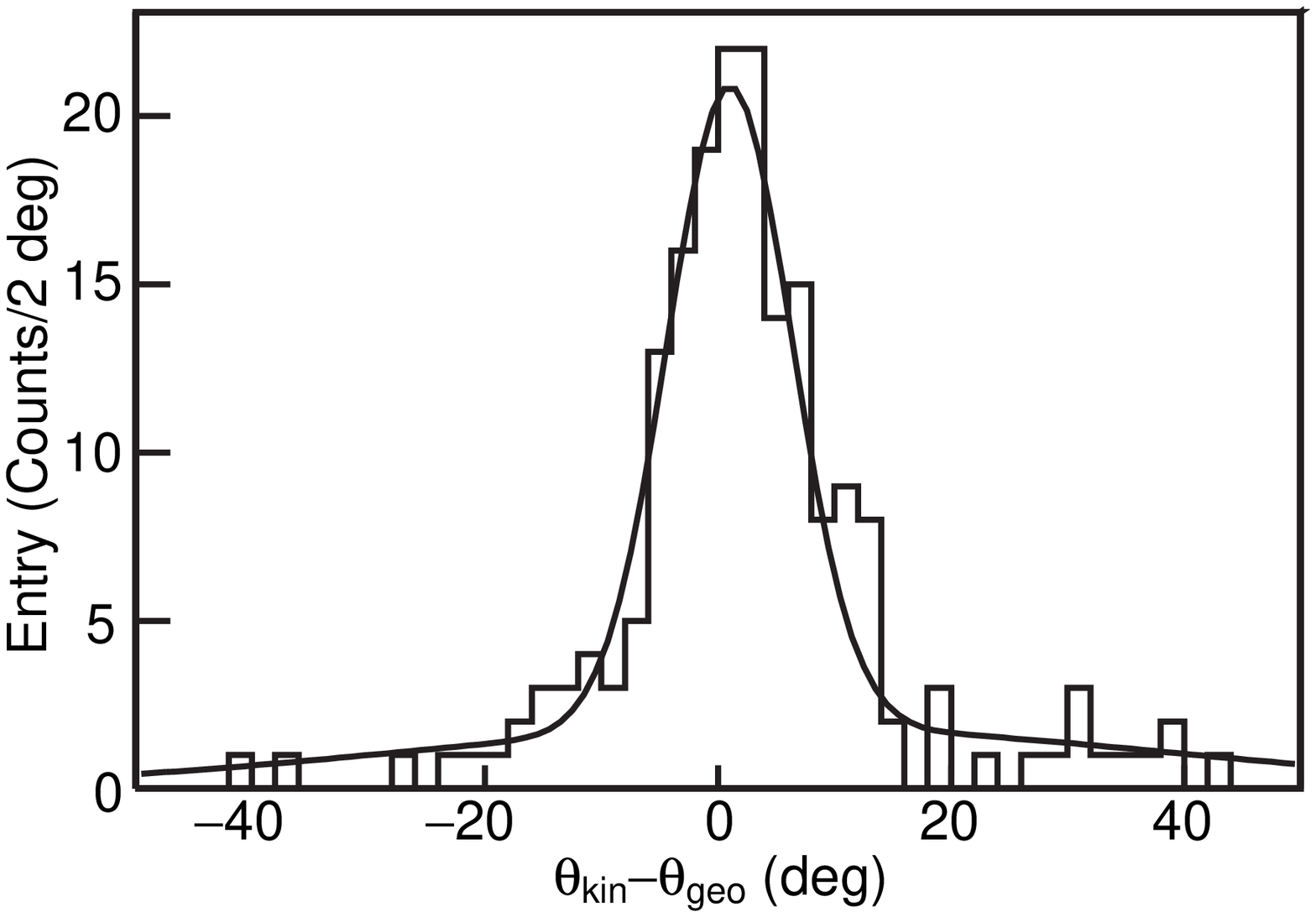}\hspace*{0.75cm} &
   \includegraphics[height=4.0cm]{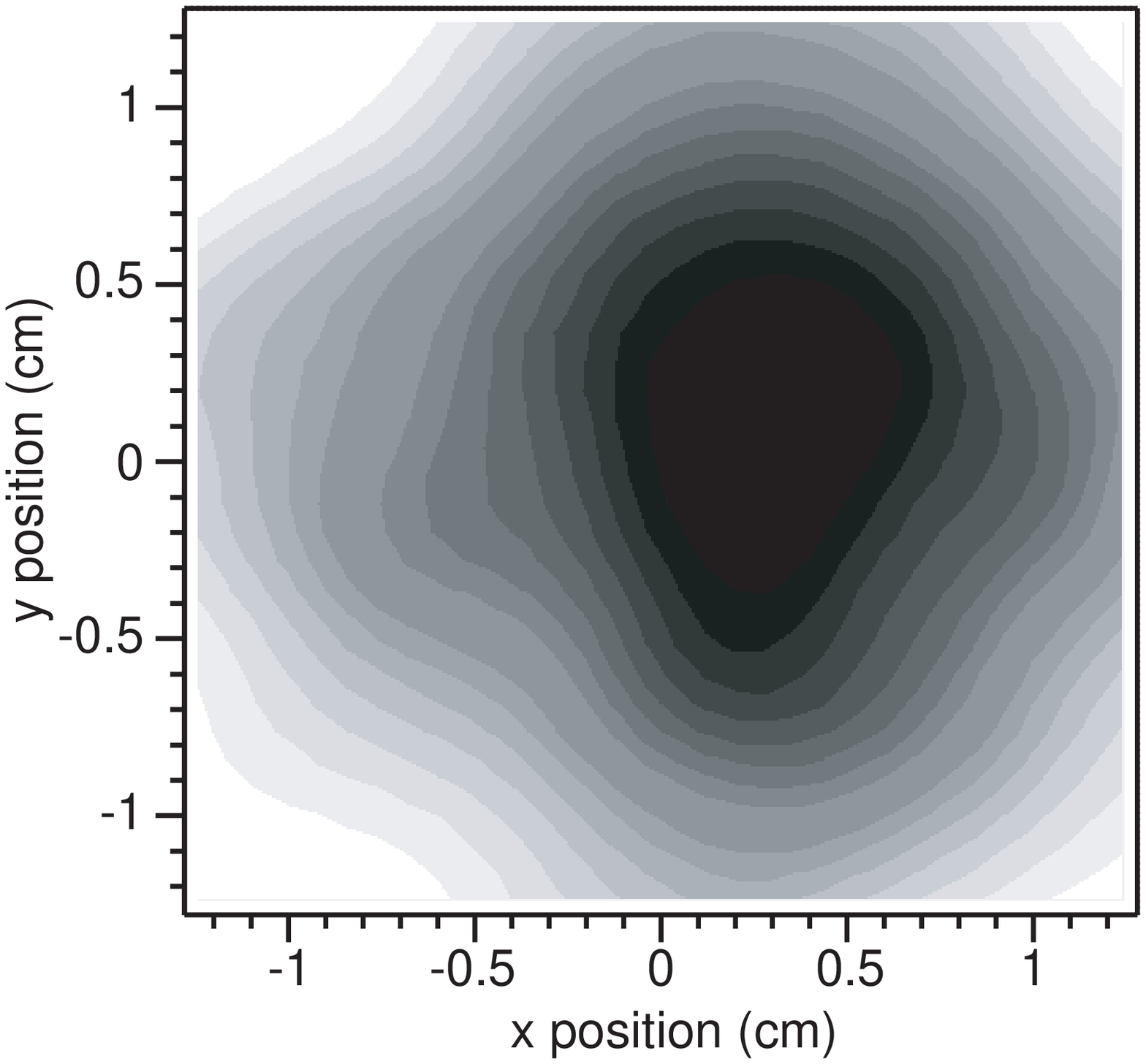}
   \end{tabular}
   \end{center}
   \caption
   { \label{fig:imaging} (a) Illustration of one Compton scattering and one photo-absorption in a single DSSD. 
(b) ${}^{57}$Co spectrum with and without background suppression using source location constraints.
(c) $\theta_\mathrm{kin}-\theta_\mathrm{geo}$ distribution for the events within 6~keV of the 122~keV peak. 
(d) ${}^{57}$Co-source image reconstructed using Compton kinematics.}
   \end{figure}

In order to investigate the $\gamma$-ray imaging capabilities, we have assembled a
DSSD module with an 800~\micron\ pitch DSSD, RC chips and VA32TA ASICs.
A ${}^{57}$Co source is placed 4~cm above the DSSD module.  
We select events in which one Compton scattering and one photoelectric absorption take place in a single DSSD as illustrated in Fig.~\ref{fig:imaging} (a).
Events with two hits on both sides and whose energy is greater than 10~keV are selected.  
Hits with larger energy deposits on both sides are paired to form a 2D hit.  
Remaining hits on both sides are paired to form another 2D hit.
The energy difference of hits in each pair is required to be less than 7~keV to suppress wrong combinations.
%The energies of two pairs are well separated for signals due to Compton kinematics as shown in Fig.~\ref{fig:imaging} (b).
The pair of 2D hits must be separated by more than 0.4~cm to ensure adequate angular resolution for the scattered photon direction. 
The open histogram in Fig.~\ref{fig:imaging} (b) shows the resulting distribution of the sum of average energies of two 2D hits.
Fig.~\ref{fig:imaging} (c) shows a $\theta_\mathrm{kin}-\theta_\mathrm{geo}$ distribution for the events within 6~keV of the 122~keV peak, where $\theta_\mathrm{kin}$ is the Compton scattering angle calculated from two energy deposits using eq.~(\ref{eq:kinematics}) and $\theta_\mathrm{geo}$ is the angle calculated from the known source location and observed positions of the energy deposits.
A fit to a sum of two Gaussian functions yields an angular resolution of 13.4\degree\ (FWHM), which is in a good agreement with the MC result of 11.8\degree.
Dominant contributions are estimated to be Doppler broadening (8.8\degree),  the energy resolution (6.0\degree) and the angular resolution of scattered photon (5.1\degree).
The latter two contributions can be reduced by using smaller strip pitch and removing RC chips.
A hatched histogram in Fig.~\ref{fig:imaging} (a) shows the energy sum distribution for the events with $|\theta_\mathrm{kin}-\theta_\mathrm{geo}|<12$\degree, which shows usefulness of such constraint.
Fig.~\ref{fig:imaging} (d) shows an image of the ${}^{57}$Co source produced by a superposition of all cones from the events within 6~keV of the 122~keV peak without a constraint on $|\theta_\mathrm{kin}-\theta_\mathrm{geo}|$.
%-------------
%-------------

In conclusion, we have fabricated prototype modules for a DSSD system,
which is a crucial element of the SCMT.  Intrinsic noise performance
is measured to be 1.0~keV (FWHM) at 0\degC\ in the DC configuration,
which is in good agreement with the analytically calculated noise
value of 0.9~keV.  The energy resolution is measured to be 1.3~keV
(FWHM) in the same configuration, indicating that an energy
resolution of 1~keV is within reach.  Gamma-ray imaging with a
DSSD using Compton kinematics is demonstrated for energy around 0.1 MeV.

\bibliography{mybib} 
\bibliographystyle{elsart-num}   %>>>> makes bibtex use mybib.bst

\end{document}